# MANY-FIELD THEORY FOR CRYSTALS CONTAINING PARTICLES WITH ROTATIONAL DEGREES OF FREEDOM

S. V. DMITRIEV[a], A. A. VASILIEV[b], A. E. MIROSHNICHENKO[c], T. SHIGENARI[d], Y. LIU[e], Y. KAGAWA[a], AND Y. ISHIBASHI[f]

[a]*Institute of Industrial Science, The University of Tokyo, Komaba 4-6-1, Meguro-ku, Tokyo 153-8505, Japan*
[b]*Department of Mathematical Modeling, Tver State University, 33 Zhelyabov Street, 170000 Tver, Russia*
[c]*Max-Planck-Institut fuer Physik komplexer Systeme, Nothnitzer Strasse 38, D-01187 Dresden, Germany*
[d]*Department of Applied Physics and Chemistry, University of Electro-Communications, Chofu-shi, Tokyo 182-8585, Japan*
[e]*National Institute for Materials Science, 1-2-1 Sengen, Tsukuba-shi, Ibaraki 305-0047, Japan*
[f]*Faculty of Communications, Aichi Shukutoku University, Nagakute-Katahira 9, Nagakute-cho, Aichi Prefecture 480-1197, Japan*

We give a brief review of some generalized continuum theories applied to the crystals with complicated microscopic structure. Three different ways of generalization of the classical elasticity theory are discussed. One is the high-gradient theory, another is the micropolar type theory and the third one is the many-field theory. The importance of the first two types of theories has already been established, while the theory of the third type still has to be developed. With the use of 1D and 2D examples we show for each of these theories where they can be and should be applied, separately or in a combination.

*Keywords:* Microscopic crystal model; rotational degree of freedom; micropolar media; continuum approximation; many-field theory

## 1. INTRODUCTION

Continuum approach gives a great simplification in many problems of mechanics and physics of crystalline and non-crystalline bodies. An important problem of continuum physics is to take into account the most essential information about the microscopic structure of matter. We will focus on the problems related to crystalline solids which have translational periodicity. The basic idea here is to start from a primitive (minimum volume) periodic element of the structure, then define the most important degrees of freedom in it, and finally, to formulate a continuum theory in terms of these degrees of freedom under the assumption that they vary slowly in space.

Simplest possible elasticity theory with three degrees of freedom for an infinitesimally small volume can be derived under the assumption that each primitive cell has three degrees of freedom, that is, components of displacement vector of its center of mass. This theory can be generalized by taking into account higher order gradient terms (the motivation is given later) [1-8]. For crystals with complicated microscopic structure the three unknown displacement fields can be not sufficient and one has to introduce additional unknown fields, e.g., fields of microscopic rotations. Generalized theories developed for these media are often called Cosserat media or micropolar media [9-18]. For the micropolar media, a strong coupling of long and short waves takes place very often. To describe this effect one needs a theory capable of description of not only long but also short waves. This theory can be constructed starting from *more than one* primitive translational cells. As for the discrete system, consideration of elementary volume with more than one primitive cell adds no new physical details. However, continuum analog for the discrete system with extra degrees of freedom is valid for both long and short waves. The theory of this type is called many-field theory [19-22].

The goal of this paper is to review the three possible generalizations of classical elasticity theory mentioned above, namely, the high-gradient theory (HGT), the micropolar theory (MPT), and the many-field theory (MFT). We will use 1D and 2D examples to demonstrate the essential features of each theory.

## 2. HIGH-GRADIENT CONTINUUM THEORY

Classical continuum theories assume that the stress in a material point depend only on the first-order derivative of the displacements, i.e. on the strains, and not on higher-order displacement derivatives. This limitation leads, for example, to unrealistic singularities near the crystal defects (in static analysis) or near the propagating shock wave (in dynamic analysis). Incorporation of the higher-order strain gradients results in regularization or smoothing of singularities or discontinuities in the strain field [6].

We consider the classical 1D chain of pointwise particles bound to the nearest neighbors. Without the loss in generality we set the mass of particles and the elastic constant of the bonds equal to unity. Then, the set of equations of motion reads

$$\ddot{u}_n = \frac{1}{h^2}(u_{n-1} - 2u_n + u_{n+1}), \qquad (1)$$

where the inter-particle spacing $h$ is the only parameter of the model, the degree of discreteness.

To derive a continuum theory one introduces the unknown displacement field $u(x,t)$ and expresses the displacements of nearest particles by means of Taylor series as

$$u(x \pm h) = u(x) \pm h u_x(x) + \frac{h^2}{2} u_{xx}(x) \pm \frac{h^3}{6} u_{xxx}(x) + \frac{h^4}{24} u_{xxxx}(x) + \ldots, \qquad (2)$$

Then, (1) can be presented in the following form

$$u_{tt} = u_{xx} + \frac{h^2}{12} u_{xxxx} + \ldots, \qquad (3)$$

When the kinematic relation $\varepsilon = u_x$ is used and the equation of motion of the continuum is expressed as $u_{tt} = \sigma_x$, the constitutive relation reads

$$\sigma = \varepsilon + \frac{h^2}{12} \varepsilon_{xx} + \ldots, \qquad (4)$$

If in the right-hand side of (4) only the first term is retained, one has classical elasticity theory, if two terms are retained, one has the second-order gradient theory, and so on.

We would like to note that the microscopic parameter $h$ is absent in the classical elasticity but it enters HGT. In this sense, HGT establishes a link between micro-scale and macro-scale.

It is instructive to compare the dispersion relations for discrete model (1) with that for classical and HGT theories. For the discrete model,

$$\omega^2 = \frac{4}{h^2} \sin^2\left(\frac{kh}{2}\right), \qquad (5)$$

where $\omega$ is the angular frequency of a phonon with the wave number $k$. For the continuum models, in general,

$$\omega^2 = k^2 - \frac{k^4 h^2}{12} + \frac{k^6 h^4}{360} + \ldots, \qquad (6)$$

which is actually the Taylor series expansion of (5) in the vicinity of $k=0$. Thus, the HGT describes the long wave part of the crystal spectrum and the accuracy of the description increases with the number of terms retained in the expansion. The classical theory has no wave dispersion because only the first term under radical is retained in this case, which gives $\omega=|k|$.

From (6), one serious problem of the second-order gradient theory can be seen. In this case, the series under the radical is truncated after the second term and thus, for $k > \sqrt{12}/h$, frequency becomes imaginary. For the continuum media the wave number is not limited by the first Brillouin zone like in the discrete model and thus, the second-order gradient theory is always unstable. To eliminate this disadvantage one has to retain the fourth-order gradient term, then the model becomes stable for any $h$.

Comparison of 1D high-order gradient models can be found, for example, in [6]. Their numerical results confirm that for the second-gradient theory the stability and uniqueness is not guaranteed and local static perturbation leads to oscillations in the entire domain. The fourth-gradient model is unconditionally stable and local perturbations remain local.

In the shock wave dynamics, the second-gradient model fails completely due to the instability of the waves with large wave numbers. The four-gradient model gives a more realistic result. Intensity of the dispersion depends on the microscopic length scale parameter $h$. However, waves with large wave numbers propagate with an unrealistically high velocity because the four-gradient model does not attempt to improve the dispersion relation for the large wave numbers.

3. FINITE SIZE PARTICLES. MICROPOLAR CONTINUUM

There are classes of materials where the displacements of the primitive cell center of mass are insufficient in description of the primitive cell state and one has to introduce additional degrees of freedom reflecting the peculiarity of microscopic structure of matter. The examples are mechanics of granular and feasible media [23], fabric mechanics [24], mechanics of periodic structures and composite materials [16], physics and mechanics of dielectric crystals [9-18,25,26], and others.

In many dielectric crystals atoms are joined in comparatively rigid clusters [27] and quartz ($SiO_2$) is perhaps the most brilliant example. In quartz, SiO$_4$ tetrahedronal clusters are almost rigid, so that β-incommensurate-α transition, observed on cooling at about 850K, occurs by mutual *rotation* of clusters with almost no deformation. Transition occurs by the mode softening mechanism so that, near the transition point, phonon spectrum has both very high frequencies, corresponding to intra-cluster vibrations, and very low frequency of the soft mode. Molecular dynamical study of this transition is a challenge because equations of motion must be integrated with a time step which is a small fraction of the shortest period of oscillation while the mode of primary interest has a very long period. The problem becomes tractable if one assumes that very hard tetrahedronal clusters are absolutely rigid. As a consequence, the highest frequencies disappear from the spectrum and the number of degrees of freedom decreases. Crystal models with rigid particles having rotational degrees of freedom have been discussed in [12,13,15,17,18,25,26]. It was demonstrated that coupling of translational and rotational degrees of freedom may result in many new physical effects: appearance of the soft optic modes, unusual elastic properties (negative Poisson ratio) [25,28-30], it may also explain many features of incommensurate phase [26,27] and others.

For the reasons given above, the importance of models with finite size particles is clear. The 1D elastically hinged molecule (EHM) model with equation of motion

$$m\ddot{u}_n = \frac{f}{h^2}(u_{n-2} - 4u_{n-1} + 6u_n - 4u_{n+1} + u_{n+2})$$
$$+ \frac{p}{h}(u_{n-1} - 2u_n + u_{n+1}) + \alpha_0 u_n + \alpha_3 u_n^3 \qquad (7)$$

is a simplest variant of such models [26]. The EHM model is a chain of rigid particles of length $h$, mass $m$, joined to each other by the elastic hinges with the elastic constant $f>0$ and subjected to the compressive external force $p$, acting along the chain. Each hinge of the chain experiences the action of the fourth-order polynomial background potential. The role of elastic hinges $f$ is to keep the chain of molecules as a straight line, while the compressive force $p$ plays a destructive role. Competition of two these parameters gives rise to modulational instability and the anharmonic background potential supports the stability of various modulated phases [31,32]. A continuum theory for EHM should take into account the properties of the carrying commensurate structure. If the carrying structure is the trivial solution, $u_n=0$, then the continuum analog to (7) reads

$$\frac{m}{h}u_{tt} = \left(fh + p\frac{h^2}{12}\right)u_{xxxx} + pu_{xx} + \frac{\alpha_0}{h}u + \frac{\alpha_3}{h}u^3. \tag{8}$$

One can see that for the crystal with microscopic rotations, the high-gradient term $u_{xxxx}$ naturally appears in the continuum theory and microscopic scale length $h$ enters the theory. In the absence of the background potential, (8) is similar to the two-gradient theory (3) with one important difference: (8) can be stable or unstable depending on the relation between $f$ and $p$, while (3) is always unstable.

The EHM model has been studied in a series of our works [26,31-35]. It has been successfully used for investigation of IC phase [35]. In the simplest 1D version, the EHM model is mathematically identical to the DIFFOUR model [36], to the model with next-nearest interaction and also the EHM model shows static properties same as the Janssen's translational invariant model [37] while the dynamic properties of these two models are different (see e.g. [32]). Continuum approximation for the EHM model enabled us to describe the transition from commensurate to a modulated phase, dynamics of domain walls [26,31], a mechanism of transition between modulated phases with different modulation wave vectors [34], and mechanisms of the lock-in transition [26,38].

The 2D version of EHM model is even richer. It contains all the physical properties of the 1D model and also explains the nature of negative Poisson ratio observed for KDP crystals and for $\alpha$-quarts near the $\alpha$–$\beta$ transition [25,30,39].

It has been demonstrated that the 2D EHM model in the long-wave approximation leads to the MPT [25]. Conditions when the micropolar elasticity equations can be reduced to the equations of conventional elasticity theory have been discussed [25].

## 4. MANY-FIELD THEORY

### 4.1 1D many-field theory

One important effect which can naturally appear in the crystals with microscopic rotations is a strong coupling between long and short waves. In the MPT derived in the long-wave limit, many important physical effects, related to the coupling, are lost.

A better continuum theory can be derived starting from more than one primitive translational cells. As for the discrete system, consideration of elementary volume with more than one primitive cell adds no new physical details. However, continuum analog for the discrete system with extra degrees of freedom is valid for not only long but also for short wave limit. This idea has been employed by Il'iushina [19,20], who studied dynamics of a simple lattice on the bases of a macro-cell containing several primitive cells. Later this idea was employed for solution of one-dimensional problems for bodies with microstructure, where the coupling takes place and the power of the method is very clear. For example, the short wavelength end-effect for a periodic frame structure has been described [21] and the buckling phenomenon for the periodically strengthened cylindrical thin shell under external pressure has been investigated [22]. Recently a many-field domain wall solution for a ferroelectric crystal has been given in [26] and compared with a single-field solution in [31]. It was demonstrated that there is a region of parameters where many-field domain wall solution can be significantly improved by retaining high-order gradient terms [31]. This is an example where MFT and HGT should be used in combination.

4.2 2D many-field theory

Let us derive the MFT for the 2D EHM model of $KH_2PO_4$ (KDP) crystal, studied in our recent paper [25]. The model consists of absolutely rigid elastically bound square particles (see Fig. 1). Elastic bonds with coefficient $C_1$ connect the vertices of each particle with the vertices of nearest neighbors. Elastic bonds with coefficient $C_2$ connect the center of each particle with the centers of next-nearest neighbors. Each particle experiences the action of the rotational background potential with coefficient $C$, has mass $M$ and moment of inertia $J$. One particle has three degrees of freedom, two components of displacement vector, *u, v*, and the angle of rotation $\varphi$.

The geometry of the model can be described by two parameters, the lattice spacing, *h*, and the parameter

$$A = \frac{\sqrt{2}}{2} a \sin \alpha, \qquad (9)$$

where *a* and $\alpha$ are the size and the orientation angle of particles, respectively.

We number particles so that *(m,n)*th particle has coordinates *(hm,hn)*. A translational cell of the model contains two particles. In [25], with the help of new variables $\phi_{m,n}=(-1)^{m+n}\varphi_{m,n}$, we used a unit cell with one particle defined by the translation vectors *(h,0), (0,h)*. Here we do not introduce new variables and use a unit cell containing two particles defined by the translation vectors *(2h,0)*, *(h,h)*. In the present case, each particle has the third index, which is 0 if the sum *m+n* is even and it is 1 if the sum is odd.

There are six degrees of freedom per one unit cell, $u_0^{m,n}$, $v_0^{m,n}$, $\varphi_0^{m,n}$, $u_1^{m,n}$, $v_1^{m,n}$, $\varphi_1^{m,n}$. Equations of motion for particle *(m,n,0)* read

$$M\ddot{u}_0^{m,n} = C_1(h^2 \Delta_{xx} u_1^{m,n} + 2u_1^{m,n} - 2u_0^{m,n}) + 2hC_1 A \Delta_x \varphi_1^{m,n} \\ + h^2 C_2(\Delta u_0^{m,n} + 2\Delta_{xy} v_0^{m,n}), \quad (10)$$

$$M\ddot{v}_0^{m,n} = C_1(h^2 \Delta_{yy} v_1^{m,n} + 2v_1^{m,n} - 2v_0^{m,n}) + 2hC_1 A \Delta_y \varphi_1^{m,n} \\ + h^2 C_2(2\Delta_{xy} u_0^{m,n} + \Delta v_0^{m,n}), \quad (11)$$

$$J\ddot{\varphi}_0^{m,n} = C_1 A^2 (h^2 \Delta_{xx} \varphi_1^{m,n} + h^2 \Delta_{yy} \varphi_1^{m,n} + 4\varphi_1^{m,n} - 4\varphi_0^{m,n}) \\ + 2hC_1 A(\Delta_x u_1^{m,n} + \Delta_y v_1^{m,n}) - C\varphi_0^{m,n}, \quad (12)$$

and equations for particle *(m+1,n,1)* read

$$M\ddot{u}_1^{m+1,n} = C_1(h^2 \Delta_{xx} u_1^{m+1,n} + 2u_0^{m+1,n} - 2u_1^{m+1,n}) - 2hC_1 A \Delta_x \varphi_0^{m+1,n} \\ + h^2 C_2(\Delta u_1^{m+1,n} + 2\Delta_{xy} v_1^{m+1,n}), \quad (13)$$

$$M\ddot{v}_1^{m+1,n} = C_1(h^2 \Delta_{yy} v_0^{m+1,n} + 2v_0^{m+1,n} - 2v_1^{m+1,n}) - 2hC_1 A \Delta_y \varphi_0^{m+1,n} \\ + h^2 C_2(2\Delta_{xy} u_1^{m+1,n} + \Delta v_1^{m+1,n}), \quad (14)$$

$$J\ddot{\varphi}_1^{m+1,n} = C_1 A^2 (h^2 \Delta \varphi_0^{m+1,n} + 4\varphi_0^{m+1,n} - 4\varphi_1^{m+1,n}) \\ - 2hC_1 A(\Delta_x u_0^{m+1,n} + \Delta_y v_0^{m+1,n}) - C\varphi_1^{m+1,n}, \quad (15)$$

where we have introduced the following differences

$$\begin{aligned}
h^2 \Delta_{xx} w_{m,n} &= w_{m+1,n} - 2w_{m,n} + w_{m-1,n}, \\
h^2 \Delta_{yy} w_{m,n} &= w_{m,n+1} - 2w_{m,n} + w_{m,n+1}, \\
2h^2 \Delta w_{m,n} &= w_{m+1,n+1} + w_{m-1,n-1} + w_{m+1,n-1} + w_{m-1,n+1} - 4w_{m,n}, \\
4h^2 \Delta_{xy} w_{m,n} &= w_{m+1,n+1} + w_{m-1,n-1} - w_{m+1,n-1} - w_{m-1,n+1}, \\
2h\Delta_x w_{m,n} &= w_{m+1,n} - w_{m-1,n}, \\
2h\Delta_y w_{m,n} &= w_{m,n+1} - w_{m,n-1}.
\end{aligned} \quad (16)$$

Dispersion relation for the model has been derived in [25]. Increase of number of particles in a unit cell from one to two means the reduction of area of the first Brillouin zone by one half and folding of the dispersion surfaces. Symmetry consideration suggests that folding occurs with respect to lines $K_x \pm K_y = \pm\pi$.

Instead of discrete variables $u_0^{m,n}$, $v_0^{m,n}$, $\varphi_0^{m,n}$, $u_1^{m,n}$, $v_1^{m,n}$, and $\varphi_1^{m,n}$, we introduce six field functions $u_0(x,y,t)$, $v_0(x,y,t)$, $\varphi_0(x,y,t)$, $u_1(x,y,t)$, $v_1(x,y,t)$, and $\varphi_1(x,y,t)$. So we use two vector functions, with indices 0 and 1, for construction of the two-field model. Continuum equations of motion with respect to these functions can be obtained from difference equations (10-15) by replacing difference operators with differential operators. Retaining the terms up to the second order we come to the long wave approximation model. In our case, in terms of new variables $U_0=u_1+u_0$, $V_0=v_1+v_0$, $\Phi_0=\varphi_1-\varphi_0$, $U_1=u_1-u_0$, $V_1=v_1-v_0$, $\Phi_1=\varphi_1+\varphi_0$, six continuum equations split into two independent groups.

Equations for functions $U_0$, $V_0$, $\Phi_0$,

$$MU_{0,tt} = h^2 C_1 U_{0,xx} + 2hC_1 A \Phi_{0,x} + h^2 C_2 (U_{0,xx} + U_{0,yy} + 2V_{0,xy}), \qquad (17)$$

$$MV_{0,tt} = h^2 C_1 V_{0,yy} + 2hC_1 A \Phi_{0,y} + h^2 C_2 (V_{0,xx} + V_{0,yy} + 2U_{0,xy}), \qquad (18)$$

$$J\Phi_{0,tt} = -C_1 A(h^2 \Phi_{0,xx} + h^2 \Phi_{0,yy} + 8\Phi_0) - 2hC_1 A(U_{0,x} + V_{0,y}) - C\Phi_0, \quad (19)$$

are the equations obtained in [25]. However, in the present case we have three more equations for functions $U_1$, $V_1$, $\Phi_1$,

$$MU_{1,tt} = -C_1(h^2 U_{1,xx} + 4U_1) - 2hC_1 A \Phi_{1,x} + h^2 C_2 (U_{1,xx} + U_{1,yy} + 2V_{1,xy}), \quad (20)$$

$$MV_{1,tt} = -C_1(h^2 V_{1,yy} + 4V_1) - 2hC_1 A \Phi_{1,y} + h^2 C_2 (V_{1,xx} + V_{1,yy} + 2U_{1,xy}), \quad (21)$$

$$J\Phi_{1,tt} = h^2 C_1 A(\Phi_{1,xx} + \Phi_{1,yy}) + 2hC_1 A(U_{1,x} + V_{1,y}) - C\Phi_1, \qquad (22)$$

Dispersion relation for the MFT (17-22), are nothing but second order Taylor series expansions of the folded dispersion relations of the discrete model (10-15) in the vicinity of the point $(K_x,K_y)=(0,0)$. This means that MFT (17-22) describes not only long waves but also short-wave oscillations with wave vectors near the corners of the first Brillouin zone, $(K_x,K_y)=(\pm\pi,\pm\pi)$.

Quality of a continuum theory can be judged from comparison of its dispersion relations with that of the discrete model (exact ones). In Fig. 2, we compare the dispersion relations of the discrete model (10-15) with that of the two-field micropolar model (17-22). We present the sections of the dispersion surfaces along $K_x=K_y$. Solid lines show the exact dispersion relations and dashed lines show different approximations. In (a), the long-wave approximation (single-field theory) derived in [25] is presented. The theory gives a good approximation only near the Γ-point. In (b), the single-field high-gradient theory (fourth order terms retained in Taylor expansions) is shown. One can see that the high-gradient theory extends the region of validity of the long-wave theory but gives no improvement near the zone boundary. In (c), the dispersion relation for the two-field continuum theory is presented. The two brunches are shown unfolded. Near the Γ-point the theory gives the accuracy of the long-wave theory (second order) but it also gives the second order approximation of the short wave region near the points $(K_x,K_y)=(\pm\pi,\pm\pi)$. In (d), the dispersion curves of the high-gradient (fourth order) two-field theory are presented.

Note that theory (17-22) combines two ideas: it is MPT, and, at the same time, MFT.

5. CONCLUSIONS

We have discussed the generalized continuum theories of three types, HGT, MPT, and MFT. HGT is *postulated* for smoothing of singularities near the crystal defects. MPT *appears* as a long-wave limit for discrete systems with microscopic rotations. MFT can be formally written for a discrete system with a simple structure, but it becomes especially important for the micropolar media because it allows one to take into account coupling between long and short waves, important for this class of materials.

**Acknowledgments**
A.A.V. wishes to thank the financial support from the grant I0754 of the Russian Federal Program "Integration" and A.E.M. wishes to thank the support from the Deutsche Forschungsgemeinschaft FL200/8-1.

Figure captions

Fig. 1. The 2D microscopic model of a crystal. Absolutely rigid square particles are bound elastically and each particle experiences the action of the rotational background potential. The lattice spacing is $h$, and $a$ and $\alpha$ are the size and the orientation angle of particles, respectively.

Fig. 2. Comparison of the dispersion relations of the discrete model (10-15) with that of the two-field micropolar model (17-22). The sections of the dispersion surfaces along $K_x=K_y$ is presented. Solid lines show the exact dispersion relations and dashed lines show different approximations, (a) long-wave approximation (single-field theory); (b) single-field high-gradient theory (fourth order terms retained in Taylor expansions); (c) two-field continuum theory (the two brunches are shown unfolded); (d) high-gradient (fourth order) two-field theory.

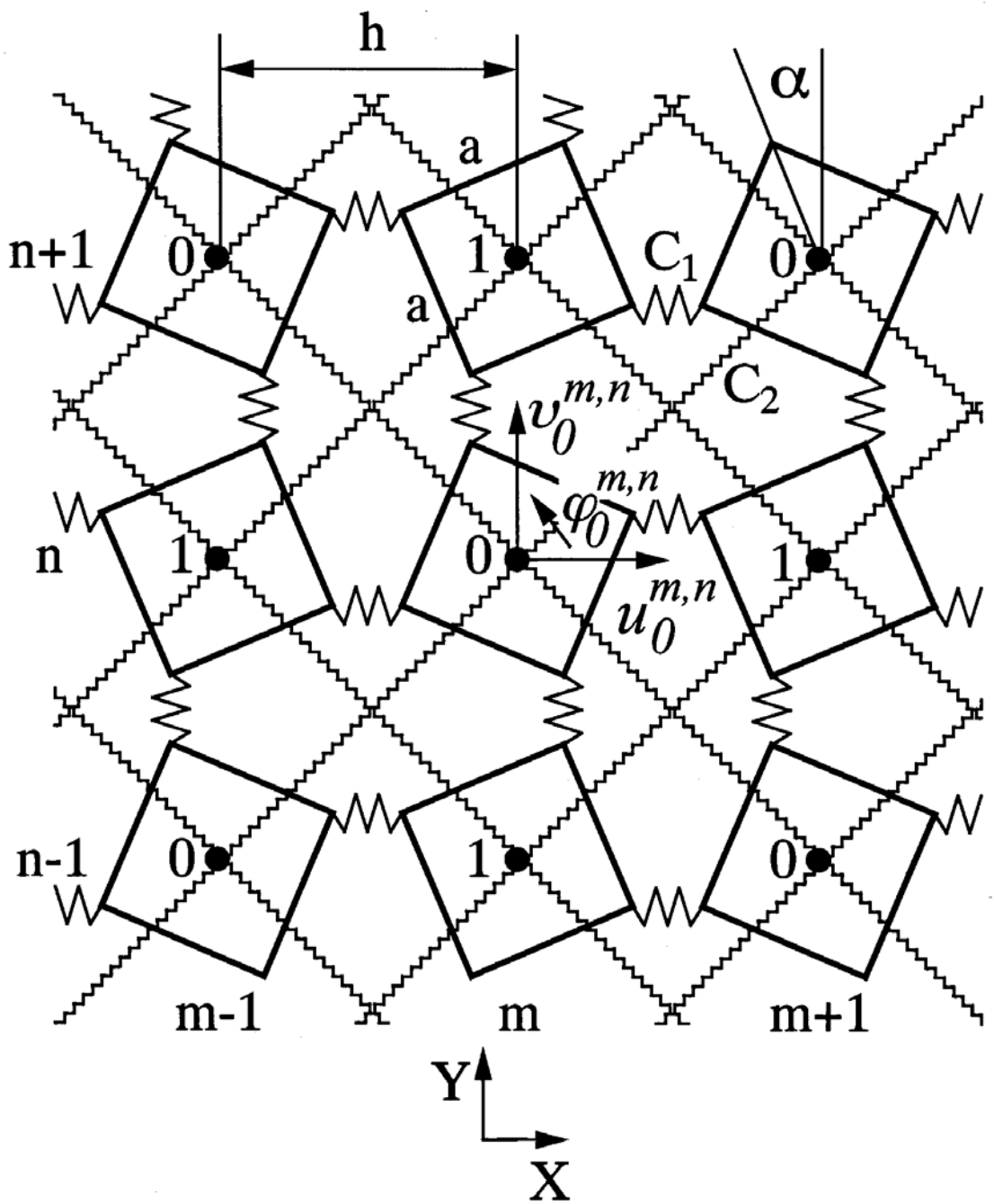

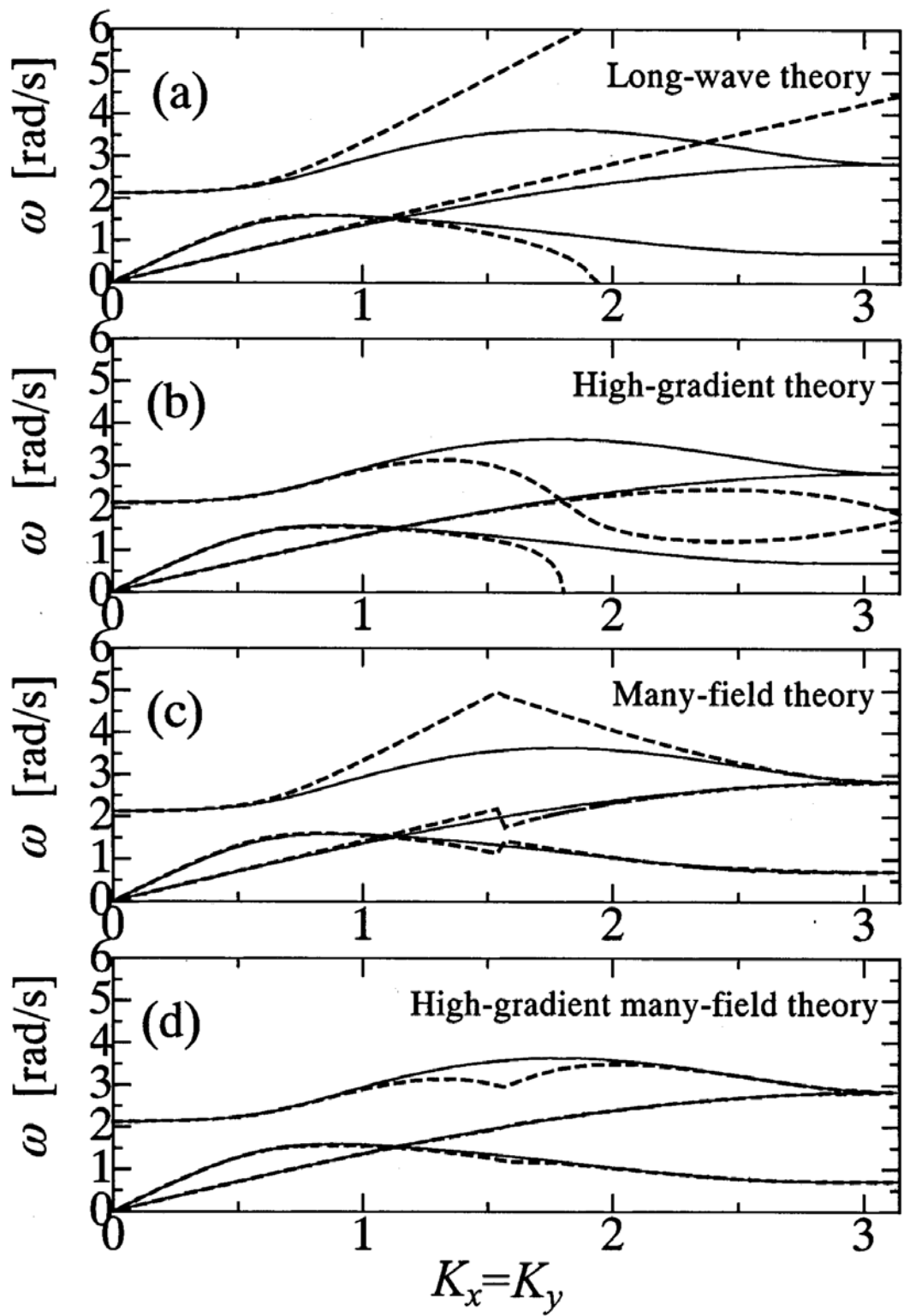